\documentclass[runningheads]{llncs}
\usepackage[T1]{fontenc}
\usepackage{graphicx}

\usepackage{blindtext}
\usepackage{pgfplots}
\pgfplotsset{compat=1.8}
\usepgfplotslibrary{statistics}

\usepackage[nolist]{acronym}  % acronym

\RequirePackage{caption}   % change figure captions
\RequirePackage{enumitem}  % change lists

\AtBeginDocument{
	
	\captionsetup{font={small}}  % mod: \sf
	\captionsetup[figure]{belowskip=0pt, aboveskip=6pt}
	\captionsetup[table]{belowskip=-0pt, aboveskip=6pt} 
	
	\setlength{\abovecaptionskip}{0pt}    %default 10pt
	\setlength{\belowcaptionskip}{0pt}         %default 0pt

}

\begin{document}
\title{Security Challenges in\\Autonomous Systems Design}

\author{Mohammad Hamad\inst{1}\orcidID{0000-0002-9049-7254} \and\\
Sebastian Steinhorst\inst{1}\orcidID{0000-0002-4096-2584}}
\authorrunning{M. Hamad et al.}
\institute{
Technical University of Munich, Munich, Germany\\
\email{firstname.lastname@tum.de}}

\maketitle              % typeset the header of the contribution
\begin{abstract}
Autonomous systems are emerging in many application domains. With the recent advancements in artificial intelligence and machine learning, sensor technology, perception algorithms and robotics, scenarios previously requiring strong human involvement can be handled by autonomous systems. With the independence from human control, cybersecurity of such systems becomes even more critical as no human intervention in case of undesired behavior is possible. In this context, this paper discusses emerging security challenges in autonomous systems design which arise in many domains such as autonomous incident response, risk assessment, data availability, systems interaction, time and data trustworthiness, updatability, access control, as well as the reliability and explainability of machine learning methods.
In all these areas, this paper thoroughly discusses the state of the art, identifies emerging security challenges and proposes research directions to address these challenges for developing secure autonomous systems.
%\TBC{The abstract should briefly summarize the contents of the paper in
%\textbf{150--250 words.}\\
%\blindtext
%}
\end{abstract}
\begin{acronym}
\acro{ECU}{Electronic Control Unit}
\acro{SOC}{Security Operations Center}
\acro{vSOC}{Vehicle Security Operation Center}
\acro{ML}{Machine Learning}
\end{acronym}

\section{Introduction}
Autonomous Cyber-Physical Systems (CPSs) are a special type of CPS where human intervention is minimal or absent. These systems are emerging in numerous application domains, including autonomous vehicles, Industry 5.0 production systems, and many others. 
Autonomous CPSs bring forth numerous benefits, enabling tasks that were formerly reliant on human intervention to be automated, enhanced adaptability in control systems, and increased safety when appropriately designed. 
Designing such systems faces many challenges \cite{seshia2016design}, including system complexity, heterogeneity, limited interpretability, and unpredictability due to Machine Learning (ML) methods. 

Among all challenges, security is probably the most critical as any security issue immediately may lead to safety problems \cite{wyglinski2013security} \cite{kim2021cybersecurity}. This is of particular importance in autonomous CPS, as there is a physical interaction with the environment, which can directly harm humans. While general information security issues can already be severe, e.g., privacy infringements, autonomous CPSs inherently entangle security and safety. In recent years, many attackers have been targeting numerous autonomous CPSs such as vehicles \cite{miller2015remote}, smart grid \cite{liang20162015}, and Industrial production \cite{dudley2021colonial}. 

The significance of security in autonomous CPSs originates from the distinct characteristics of the system itself. These characteristics include the "no-human-in-the-loop" paradigm, necessitating the system's self-awareness of potential attacks. Furthermore, security solutions must align with CPS properties, encompassing the embedded nature that restricts resource-intensive approaches. Here, the distributed and heterogeneous system nature mandates the necessity for distributed security solutions. Additionally, the system's hard real-time demands, wherein tasks and messages must adhere to predefined time limits, call for lightweight security solutions that can be efficiently computed on embedded platforms. As a consequence, the safety-criticality of autonomous CPS obligates security to be an indispensable component.

\begin{figure}
    \centering
    \includegraphics[width=\textwidth]{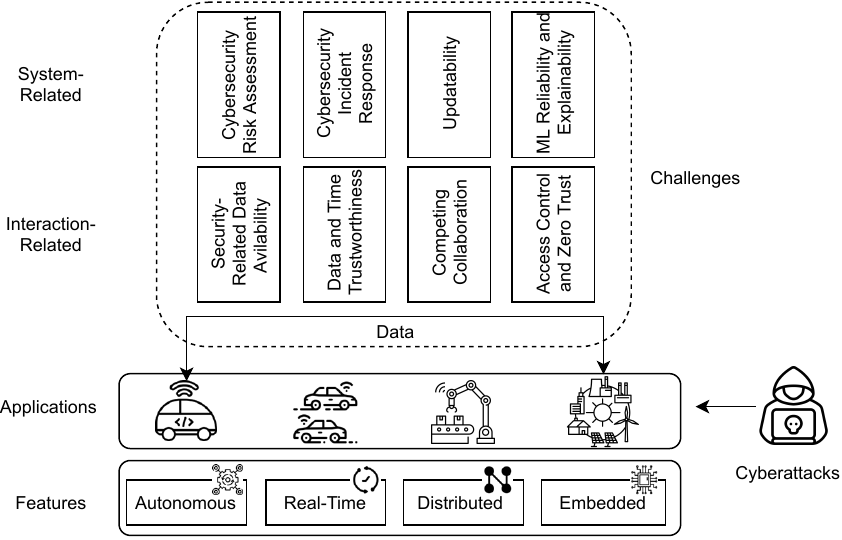}
    \caption {Security Challenges in Autonomous System Design}
    \label{fig:fig1}
\end{figure}

In this paper, we will thoroughly examine the myriad of security challenges that present significant obstacles to the realization of autonomous CPSs. These challenges encompass the system's capacity for autonomous responses to cybersecurity threats and its ability to effectively evaluate the associated risks (see Fig. \ref{fig:fig1}). Furthermore, we will delve into the issue of limited comprehensive data, which restricts the efficacy of ML-based approaches for detecting such attacks. Additionally, we will discuss in detail the hurdles related to the system's collaboration within a computing environment, all while ensuring the utmost trustworthiness of data and time during these collaborative efforts and feasible and efficient access control methods. Furthermore, we will investigate the intricate task of ensuring the continuous and autonomous updating of the system, particularly within the context of a complex and ever-evolving ecosystem. Lastly, we will discuss 
 the issue of explainability and reliability concerning the various ML models integrated into autonomous systems.  
In the remainder of the paper, we will explore these challenges in greater detail.

% initial thoughts 
%\begin{itemize}
    %\item Autonomous Systems are emerging in many application domains. (Industry 4.0 and autonomous production systems: Industry 5.0: AI + HMI )
    %\item Benefit: Performing tasks that previously required human action/interaction, increased adaptability in control systems, increased safety when designed properly...
    %\item Challenges in many domains: Complexity, limited interpretability, and predictability due to machine learning methods, ...
    %\item Among all challenges in autonomous systems design, security is probably the most critical one, as any security issue immediately may lead to safety problems.
    %\item  This is of particular importance in ACPS, as there is a physical interaction with the environment that can directly harm humans. While general information security issues can be severe regarding, e.g., privacy infringements, cyber-physical systems inherently entangle security and safety. This is applicable to all domains of autonomous CPS:  AV is one main example (say some obvious issues for motivation), 
   
    %\item what makes ACPS security critical: no human, self-aware about the attack. But it is still embedded, safety-critical, distributed 
    %\item 
%\end{itemize}

\section{Autonomous Cybersecurity Incident Response} 

In safety-critical autonomous systems, such as autonomous vehicles, the response to an attack is as crucial as detecting the attack itself. These systems must be capable of autonomously responding to attacks, eliminating the need for human intervention — especially in fully autonomous scenarios.\footnote{It is crucial to distinguish between automated cybersecurity incident response, which is a pre-programmed response triggered each time an attack is detected, and autonomous incident response, which is generated and decided by the system independently.} In the case of autonomous vehicles, fortunately, cybersecurity incident response is an integral aspect of security management, as outlined in ISO/SAE 21434 within the operational and maintenance clause \cite{international2021iso}. Based on the standard, this process aims to provide remedial actions and updates, which may involve post-development changes to address security vulnerabilities. The process necessitates the vehicle to share cybersecurity information about the vulnerability that triggered the cybersecurity incident response.

To realize such capabilities, \ac{SOC} \cite{vielberth2020security} or application specific \acp{SOC}, such as \ac{vSOC} \cite{sembera2020iso} \cite{olt2019establishing}, are utilized to facilitate monitoring and managing cybersecurity incidents. Within each \ac{SOC}, numerous human experts and/or services collaborate to analyze and address reported incidents. These \acp{SOC} are generally centralized to serve multiple systems that share incident information with them \cite{10012711} \cite{barletta2023v}. 
Nevertheless, this solution does not appear scalable or capable of delivering the near-real-time responses required for many autonomous systems, particularly within autonomous vehicles. This limitation can arise from various reasons:
\begin{itemize}
    \item \textbf{Stable Connectivity}: There is a crucial need for stable connectivity to facilitate the exchange of cybersecurity information data between the systems and the \ac{SOC}. However, establishing and maintaining such connectivity can present challenges, especially in specific areas with mobile autonomous systems or where the connectivity becomes unavailable due to a cyberattack. 
    \item \textbf{Data Volume}: The substantial volume of data shared from a large number of systems can require significant processing resources and time within the \ac{SOC}.

\item \textbf{Extended Incident Response Time}: The time required to address cybersecurity incidents might sometimes extend to days or weeks, even if the compromised component is critical \cite{european2019enisa}.

\end{itemize} 

Therefore, there is a necessity to have an \textit{local} autonomous incident response system that can seamlessly integrate with each node of the system (e.g., each autonomous vehicle). This would eliminate the need for exchanging substantial amounts of information with a \ac{SOC} and empower the node to autonomously, and near-real-time react to detected attacks without waiting for remote responses that might be impractical. However, developing such a system must meet specific requirements stemming from the unique nature of each system. 
For example, in autonomous vehicles, this system should address resource constraints, real-time processing, distributed architecture, and the interconnected nature of autonomous vehicles. Additionally, the provided system should possess the capability, supported by an autonomous and dynamic risk assessment system, to proactively and dynamically respond to various attacks \cite{mhhirs}.  
Importantly, such a system should not replace the \textit{global} \ac{SOC} but rather be an integral part of a holistic multi-layer response system where some information is still shared with the \ac{SOC} to facilitate a comprehensive system-wide response. 
%Moreover, the integration of such technology into autonomous vehicles necessitates that the vehicle's system architecture is capable of accommodating and implementing changes during runtime. This involves the system's ability to facilitate runtime reconfiguration, which may encompass features like task migration from one Electronic Control Unit (ECU) to another, dynamic alteration of security policies for various software components, and ensuring real-time constraints within the system.

\section{Autonomous and Dynamic Cybersecurity Risk Assessment}

The intrusion detection mechanism has undergone extensive research and investigation, resulting in a robust and well-documented system. However, evaluating the risk associated with detected attacks during runtime is still an emerging area of study, especially concerning the context of cyberattack risks. 
Fully autonomous systems should possess the ability to autonomously analyze data, detect potential vulnerabilities, identify threats, and evaluate associated risks. Ideally, these processes should occur with minimal or no human intervention. 
%However, the current state of the art in this field is far from achieving this goal. 
%
%In the domain of autonomous vehicles, threat analysis and risk assessment (TARA), a crucial component of ISO/SAE 21434 \cite{international2021iso}, aids Original Equipment Manufacturers (OEMs), Tier 1 suppliers, and other entities in the automotive supply chain to comprehend and prioritize potential threats and risks across various sub-systems and components of the automotive system. TARA assists in effectively allocating resources and shaping security strategies, policies, procedures, and controls. 
The risk assessment process is typically employed to enable the system owner to understand and prioritize potential threats and risks across various sub-systems and components of the autonomous system. 
%TARA assists in efficiently allocating resources and shaping security strategies, policies, procedures, and controls.
This process mostly takes place during the design phase, influencing decision-making and guiding security-related development.\footnote{ It's important to recall that risk management encompasses the entirety of a component's lifecycle, whereas risk assessment serves as a distinct element within the broader scope of risk management. } 
The current risk assessment process suffers from the following issues: 
\begin{itemize}

    \item \textbf{Emergent Runtime Risks}: 
As previously mentioned, the process of risk assessment primarily occurs during the design phase and heavily relies on the information available at that time. Consequently, if a new attack arises during runtime that was not accounted for during the assessment phase (due to incomplete information or a zero-day attack), the accuracy of the assessment results could be compromised.
    \item \textbf{Runtime Context Unawareness}:  In many domains, the risk assessment process neglects the system's runtime context, operational status, and the surrounding environment. This oversight can be attributed to a lack of comprehensive understanding of damage scenarios during various system operational states. It is crucial to differentiate between risks arising from attacks during different system operational states and how these states affect the computed risk.
For instance, in the domain of autonomous vehicles, attacks may target the vehicle while it is driving, parked, or undergoing maintenance. However, the vehicle's status should influence the assessed risk. Compromising the GPS while the vehicle is parked in a parking lot versus during highway driving should result in different risk levels. While damage scenarios can partially address this issue in the context of autonomous vehicle risk assessments, it is important to emphasize the diverse impact of attacks across various operational states of the vehicle, as well as other factors like weather conditions, road conditions, time of day, and more.
 \item \textbf{Safety and Privacy Overlooks Security}: The current risk assessment process gauges the damaging impact of attacks primarily based on safety, financial, operational, and privacy factors. However, this approach tends to overemphasize safety parameters, which is justifiable considering the safety-critical nature of intelligent vehicles and their financial significance to system owners and users. Even if privacy considerations are included, this approach still falls short of encompassing the full spectrum of cybersecurity attacks. It's essential to recognize that assuming safety and privacy are the sole impacts of attacks is not entirely accurate. While cybersecurity attacks can potentially compromise the functionality of autonomous systems and data, not all of them will directly lead to safety hazards or privacy concerns.
\end{itemize}

All these issues should be addressed by a comprehensive risk assessment method. 
The aim of this risk assessment method should surpass the current objective of solely aiding the decision-making process for choosing security measures during the development of autonomous vehicle systems. Instead, it should empower the system to make instantaneous decisions in the face of cybersecurity attacks while the system is operational \cite{vehits19}.
Furthermore, this method should seamlessly operate without human intervention. It is possible to provide valuable assistance in this area by using \ac{ML} approaches \cite{paltrinieri2019learning}. Furthermore, using dynamic safety risk assessment methods (e.g., \cite{patel2021machine} \cite{9462019}) is essential for enhancing effectiveness.

\section{Security-Related Data Availability}
\label{sec:data}
One of the main characteristics of autonomous CPSs is their ability to learn and adapt, typically accomplished through the extensive use of ML-based solutions. For instance, in the realm of autonomous vehicles currently navigating streets, ML algorithms are already employed to enhance advanced driver-assistance systems (ADAS) \cite{walz2020waymo}. With \ac{ML}'s expansion, the technology will handle complex tasks, including those related to cybersecurity. 
Leveraging ML brings numerous benefits, particularly its capacity to process vast data volumes from sensors, self-adapt and evolve, and accuracy when trained on high-quality, pertinent, and representative data. 

While numerous datasets are available for training \ac{ML} models, there remains a demand for additional data \cite{paleyes2022challenges}.  
For instance, in the domain of autonomous vehicles, the available datasets are often limited to just one or two sensors and may not represent diverse real-world scenarios, resulting in datasets that do not accurately reflect real-life outcomes. This issue becomes even more pronounced when training \ac{ML} models are designed to detect cybersecurity attacks. Such models require not only nominal datasets but also datasets containing off-nominal data that simulate attacks on the system. Unfortunately, the availability of such comprehensive datasets suitable for testing and validating ML-based intrusion detection mechanisms is quite limited and frequently lacks detailed attack information \cite{he2020machine} \cite{braunegg2020apricot}. 
Generating realistic attack data by exploiting actual vulnerabilities of autonomous systems in an operational environment is a highly challenging process. This challenge arises not only due to technical difficulties but also due to various constraints, including confidentiality rights, vulnerability disclosure, hardware costs, and more. Moreover, attacks on safety-critical autonomous systems face additional hindrances stemming from regulatory and legal requirements \cite{winner2018introducing}.

An optimal approach to generating such datasets might involve utilizing digital twins and simulation-based technologies instead of actual vehicles \cite{10200555}\cite{piromalis2022digital}. This strategy can establish a secure and effective environment for illustrating the repercussions of diverse cyberattacks on autonomous systems, enabling the testing, validation, and assessment of security mechanisms' efficacy against potential attacks. While initial steps have been taken toward achieving this goal in the domain of autonomous vehicles, some solutions are constrained to specific smart sensors \cite{9774542}, \ac{ML} models \cite{nesti2022evaluating}, or communication aspects \cite{9908023}. More advanced options, like \cite{10200555}, cover numerous sensors and communication modalities, but the development of comprehensive solutions encompassing all sensors and facilitating sophisticated attacks remains limited. Additionally, the validation of datasets generated by such frameworks poses a significant challenge that necessitates resolution. 

\section { Interaction of Autonomous Systems }
The interaction of several autonomous systems in a system of systems, such as autonomous vehicles exchanging perception data among each other within a certain proximity, brings additional security challenges beyond conventional communication security perspectives. There is, e.g., an ongoing discussion in the automotive community about whether future self-driving vehicles should be really fully autonomous or rather infrastructure-guided. In the case of full autonomy, each vehicle is making its own decisions, only relying on its own sensor data and hence limiting possible attack vectors. Here, each vehicle will individually optimize its goals of reaching its destination as fast as possible, maximizing passenger safety, etc. However, this can lead to competitive behavior, where ML algorithms utilize legal but, from a human driver perspective, unethical behavior that may help to, e.g., navigate a traffic jam faster than other vehicles by triggering their safety routines due to reducing the distance to other vehicles. 

In the case of future cooperative and collaborative autonomous systems where information is exchanged for the benefit of all participants, such behavior may lead to a class of autonomous systems, which we introduce as \emph{Competing Collaborative Systems} (CCS). Such CCS may emerge in industrial production systems where several machines in a factory, provided by different suppliers "as a service", will compete to get a particular manufacturing job. In CCS, security becomes multi-facetted, as adversarial behavior is no longer necessarily purely exercised by malicious participants but rather as part of the interaction of CCS. Here, we will see the challenge of the interaction of nodes in open systems where joint security agreements can only be very limited. Imagine an infrastructure-guided form of future autonomy, where sensors in the traffic infrastructure provide perception data to the vehicles. Attacking such infrastructures will lead to catastrophic results and mandates both highest-level security efforts for the interaction in the system as well as roots of trust for traffic data. Nevertheless, vehicles in the CCS context will need to perform plausibility checks of received information and may consider external information as potentially malicious until proven otherwise.

In another view on the interaction of autonomous systems, interoperability and compatibility of various security protocols and requirements, that must be enforced by different nodes, pose significant challenges. Some nodes in open systems may have security levels that others do not prefer or accept. Ensuring the system's functionality without compromising the overall security level requires a clear mechanism for exchanging information and reaching agreements among different nodes. This is exacerbated by the general requirement of integrating existing legacy systems where current security paradigms may not exist. 

From an autonomous system design perspective, the above-mentioned challenges may lead to a new class of design paradigms where trust among system nodes needs to be redefined, which we will investigate in the subsequent section.

\section{Data and Time Trustworthiness}
%\TBC{@Sebastian:   we may cite the sok paper somewhere here \\}
In many autonomous CPSs, trust between nodes is essential to facilitate collaboration and prevent collisions. Paradoxically, this reliance on trust can often lead to vulnerabilities that nodes can exploit to compromise the system \cite{7542961}. Ensuring the trustworthiness of exchanged data can be particularly challenging, especially within CCS, where each node or entity competes for privileged access to shared resources. For instance, let's consider a scenario at an intersection involving autonomous vehicles, where each vehicle strives to be the first to pass through in order to minimize traffic congestion.
The challenge becomes even more pronounced when we need to maintain the trustworthiness of information that humans may interact with within the system. This is especially pertinent when we acknowledge that humans often possess a more comprehensive understanding and better vision regarding the ongoing situation.
In such cases, if a node or a human attempts to deceive or manipulate the system for their advantage, relying solely on secure communication (which ensures data integrity and authenticity) among the nodes is insufficient to prevent these situations from arising.
Ensuring trustworthiness requires not only data integrity, authenticity, and availability but also data accountability, traceability, transparency, and validity. Achieving each of these requirements is challenging, especially when considering the high heterogeneity of most of autonomous systems and the multitude of sensors and subsystems involved in acquiring and processing information. 

As much as ensuring data trustworthiness, autonomous systems also need to ensure \textit{data time trustworthiness}, which refers to the accuracy and reliability of timestamps associated with messages exchanged between autonomous nodes. In many autonomous systems, trustworthiness of time data is critical for coordinating the interactions among various nodes. Security attacks targeting the timing associated with exchanged data can disrupt the system and may lead to catastrophic consequences \cite{sargolzaei2014delayed}. 
To ensure the trustworthiness of message/data timing, one common approach is to use time synchronization protocols such as the Precision Time Protocol (PTP) \cite{ieee-1588-2019}. PTP is widely employed in many safety-critical autonomous systems, including industrial production, autonomous vehicles, and smart grids. The security of PTP and similar protocols has been under investigation for some time, resulting in proposed security solutions that have become part of the standard, such as adopting security protocols like IPsec or MACsec to mitigate various attacks that could target these protocols.
However, a significant security issue that remains is related to time delay attacks on PTP \cite{10001666}. In such attacks, the attacker selectively delays the reception of certain protocol messages that contain timestamps used for synchronizing the node with a master node without manipulating the content of the messages. In other words, the attack targets the integrity of the timing information rather than the content of these messages. 
Maintaining trustworthy interactions among nodes in autonomous systems requires securing PTP and similar protocols against time delay attacks.

\section{Autonomous  and Continuous Updatability}
Autonomous CPSs may consist of various subsystems owned by different entities and suppliers. Take, for instance, an autonomous vehicle, which incorporates between 70 to 100 \acp{ECU} hosting over 100 Million lines of code. In the development of such a vehicle, numerous companies are typically involved. Within such complex systems, it's imperative to ensure that all components, including hardware, firmware, and software, remain up-to-date, particularly in cases where patches are needed to address security vulnerabilities. However, notable incidents like Stuxnet \cite{matrosov2010stuxnet} and the WannaCry ransomware attacks \cite{10099169} underscore the challenges of maintaining up-to-date systems. These attacks exploited unpatched vulnerabilities. The patches for these vulnerabilities were introduced in advance; however, many users did not adopt these updates. 
These examples highlight the impracticality of achieving this even when a small number of entities are responsible for the task. This raises concerns about the feasibility of keeping such systems updated, especially in scenarios where many entities need to be involved in providing updates, and no human is available to handle these updates.

Ensuring system updatability requires the existence of secure Over-the-Air (OTA) update mechanisms. Such mechanisms have already been developed or mandated in a few domains, such as autonomous vehicles \cite{halder2020secure}. However, in most cases, patching and software updates over-the-air are not feasible solutions for low-end devices, as they do not support such functionality \cite{enisa2019industry}.
Another challenge is to ensure that all components are up to date. The existence of components that use outdated software or libraries may serve as a gateway to compromise the entire system using different attacks, such as downgrade attacks. Considering the high number of entities responsible for providing security updates for different components, it is possible that one or many of them may not provide such patches for various reasons.
One of these reasons could be that the company no longer exists, without transferring the responsibility to update its components to another entity, or simply because the company has stopped supporting the old version of the software due to the release of a new version. 
To ensure continuous updatability, the autonomous system should be designed to support OTA capability, maintain upgrade dependencies, mitigate the presence of outdated components in the system (which may limit their interaction with critical components), and finally, this should be supported by standardization efforts and legal legislation to guarantee the provision of security updates for each part of the system for a particular time.

% 
% zero trust  for autonomous systems  and context-aware firewalls 

\section{Access Control and Zero Trust}
Autonomous systems are designed to function within dynamic and rapidly changing environments where contextual factors and operational requirements can evolve swiftly. This unique characteristic presents a considerable challenge to traditional security measures, rendering them ineffective in managing the complexities of these ecosystems. Factors such as location, time, and the system's status all play critical roles in determining the appropriate security measures \cite{SEEMQTT}. Security policies in autonomous systems must be equipped to accommodate these dynamic aspects. This often involves the authorization of specific actions at certain times and the restriction of the same actions at others. For instance, within the automotive domain, a vehicle's security system might permit specific actions when the vehicle is stationary in a workshop or parking lot but restrict those same actions when the vehicle is in motion. Compounding the complexity, autonomous systems comprise a vast number of nodes, each containing a significant number of components. Take, for example, an autonomous vehicle, which encompasses a multitude of components, making it infeasible to scale traditional access control mechanisms effectively. The dynamic nature of these systems and the various factors that must be considered challenge the conventional approach to access control. 

In recent years, the concept of \textit{Zero Trust} has emerged as a prominent approach in research and practical applications, particularly in response to the evolving security needs of complex systems \cite{yan2020survey}. At its core, Zero Trust challenges the traditional assumption of trust within an ecosystem. The fundamental premise is that no component or node within the autonomous system should be inherently trusted. 
To establish a zero trust framework, various technologies are required to support critical functions like user identification, authentication, and secure communication. This approach operates under the fundamental belief that trust cannot be assumed for any component or node within the autonomous system. While cryptographic solutions offer potential remedies to address these challenges, not all are suitable, primarily due to the limited capabilities of resource-constrained devices commonly integrated into autonomous systems. Careful consideration is necessary when selecting cryptographic approaches to ensure compatibility with the system's constraints and requirements.

\section{ML Reliability and Explainability}
Ensuring the reliability of \ac{ML} models is paramount, as any unreliable decision made by these models can result in catastrophic consequences. \ac{ML} reliability can be undermined by various attacks during different operational phases of the models, including both the training and online phases. 
During the training phase, the model's reliability relies heavily on the quality and availability of training data, as discussed in Section \ref{sec:data}. 
Attacks such as data poisoning \cite{9855872} can mislead the model into making targeted misclassifications or reduce model classification accuracy.
Another significant challenge lies in ensuring the robustness of these models during run time, enabling them to effectively handle adversarial inputs. 
In recent years, adversarial machine learning attacks have become a pressing concern. These attacks involve the creation of adversarial examples by introducing small, often imperceptible, changes to the input data. Such attacks have targeted numerous \ac{ML} models in autonomous systems \cite{9032989}, resulting in severe vulnerabilities and potential risks. Addressing these challenges is crucial to ensuring the safety and reliability of autonomous systems reliant on \ac{ML}. 

Another significant concern about ML models in autonomous systems is their inherent uncertainty, which can limit their use. Due to this, ensuring the explainability of ML models and providing reasoning for their decisions, especially in the context of security, becomes crucial. 
Having humans as part of an autonomous system's ecosystem and requiring interaction between the autonomous system and humans makes explainability and reasoning even more necessary \cite{chamola2023review}. The decisions made by ML models in such scenarios need to be understood and trusted by humans, especially when they have security implications. 
By enhancing explainability in ML models, more transparency and collaboration can be achieved, resulting in more trustworthy and reliable autonomous systems \cite{glomsrud2019trustworthy}. 
%\begin{itemize}
%\item As we have mentioned, ML is becoming an essential part of an autonomous system. Ensuring the reliability of the ML model is critical since any unreliable ML-based decision may lead to catastrophic consequences
%\item the reliability of the ML learning model depends on the quality and the existence of data that can be used to train the model (which is discussed in section XX) 
%\item Another challenge is ensuring the robustness of the model and its capability to handle adversarial inputs. 
%\item Finally, the ever-changing conditions around some autonomous systems require that the ML models are capable of handling the new data produced by these new conditions without manipulating the system's resilience. 
  %  \item Ensuring the explainability of ML models is essential. While ML can be valuable, it's crucial to provide reasoning for the decisions made by these models, particularly concerning security. 
  %  \item This is crucial in case a human is part of the ecosystem of the autonomous system, and there is a need for autonomous system-human interaction. 

  %  %\item  Adversary machine learning 
%\end{itemize}

\section{Conclusion}
With the fast adaption of autonomous systems in many areas of technology, critical security challenges arise which need to be identified and addressed before such systems are deployed. For this purpose, this paper discussed emerging security challenges in many domains of autonomous systems design such as autonomous incident response, risk assessment, data availability, systems interaction, trustworthiness, updatability, access control, as well as the reliability and explainability of machine learning methods. Such security aspects are traditionally not given sufficient attention in system design processes. However, in autonomous systems, the catastrophic consequences of any kind of security attack require the integration of security considerations as early as possible in the design process. Consequently, in this paper, we propose research opportunities to contribute to the development of secure autonomous systems.

\subsubsection{Acknowledgments} 
This work is supported by the European Union-funded project CyberSecDome (Agreement No.: 101120779). 

 \bibliographystyle{splncs04}
\bibliography{reference}

\end{document}